\documentclass[aps,prd,amsmath,amssymb,reprint,showpacs,superscriptaddress,nobibnotes,floatfix]{revtex4-1}
\usepackage{CDFRevTeX}

\usepackage{xspace}
\usepackage{standalone}
\usepackage{booktabs}
\usepackage{notoccite}
\usepackage{multirow}

\usepackage[tracking=true,kerning=true,expansion=true,spacing=false,factor=1100,stretch=20,shrink=20]{microtype}
\SetTracking{encoding={*}, shape=sc}{50}

\usepackage{siunitx}
\DeclareSIUnit\clight{\text{\ensuremath{c}}}

\usepackage[pdftex,unicode=true,pdfpagelabels=true,pdfusetitle,pdfauthor={The CDF Collaboration},hidelinks]{hyperref}

\usepackage{tikz}

\newcommand{\noteprefix}{CDF/ANAL/TOP/CDFR}
\newcommand{\notenumber}{11146}
\newcommand{\version}{4.0}

\newcommand{\bbbar}{\ensuremath{b\bar{b}}\xspace}
\newcommand{\ttbar}{\ensuremath{t\bar{t}}\xspace}
\newcommand{\ppbar}{\ensuremath{p\bar{p}}\xspace}
\newcommand{\bbar}{\bbbar}
\newcommand{\afb}{\ensuremath{A_\text{FB}}\xspace}
\newcommand{\afbb}{\ensuremath{A_\text{FB}(\bbbar)}\xspace}
\newcommand{\mbb}{\ensuremath{m_{\bbbar}}\xspace}
\newcommand{\abs}[1]{\ensuremath{\left\lvert #1 \right\rvert}\xspace}

\newcommand{\gevM}[1]{\SI[per-mode=symbol]{#1}{\giga\eV\per\clight\squared}\xspace}
\newcommand{\gevE}[1]{\SI{#1}{\giga\eV}\xspace}

\newcommand{\tabref}[1]{Table~\ref{#1}}
\newcommand{\figref}[1]{Fig.~\ref{#1}}
\newcommand{\Figref}[1]{Figure~\ref{#1}}
\newcommand{\secref}[1]{Sec.~\ref{#1}}
\let\origeqref\eqref
\renewcommand{\eqref}[1]{Eq.~\origeqref{#1}}
\newcommand{\Eqref}[1]{Equation~\origeqref{#1}}

\begin{document}

\title{
First measurement of the forward-backward asymmetry in bottom-quark pair production at high mass}

\date{\today}

\begin{abstract}
We measure the particle-level forward-backward production asymmetry in \bbbar pairs with masses (\mbb) larger than \gevM{150}, using events with hadronic jets and employing jet charge to distinguish $b$ from $\bar{b}$. The measurement uses \SI{9.5}{\per\femto\barn} of \ppbar collisions at a center of mass energy of \SI{1.96}{\tera\eV} recorded by the CDF II detector.  The asymmetry as a function of \mbb is consistent with zero, as well as with the predictions of the standard model.  The measurement disfavors a simple model including an axigluon with a mass of \gevM{200} whereas a model containing a heavier \gevM{345} axigluon is not excluded.
\end{abstract}


\pacs{14.65.Ha}

\maketitle

\section{Introduction}
\label{sec:theory}

In recent years, the values of the forward-backward asymmetry (\afb) of top-quark-pair production measured at the Tevatron proton-antiproton collider at Fermilab have been consistently larger~\cite{CDFLJAfb,*D0LJAfb} than those predicted by the standard model (SM)~\cite{MitovAfb,BernSi,KuhnRodrigo99,HollikEWK,KuhnEWK,ManoharEWK}.  Further study of this phenomenon has led to a number of proposed extensions of the SM~\cite{CaoRosner,*greshamkimzurek,*JungPearceWells,*BSMReviews}.  One specific class of such models is the low-mass axigluon~\cite{AguilarLightAxiG,*Schmaltz,*Falkowski,*Ipek}. These models include a new axial-vector boson with a mass below the \ttbar threshold and a natural width broad enough to evade detection in light-quark resonance searches, and predict a non-zero forward-backward asymmetry due to the interference between amplitues mediated by the gluon and the axigluon.  In the simplest models, the axigluon has equal couplings to all the quarks.  Hence, a stringent test of such models is the measurement of the forward-backward asymmetry of pair production of other quark flavors, such as bottom.

\nocite{geometry}

\begin{table}


\caption{Predicted values of \afb in the standard model~\cite{Grinstein} and in two models with an axigluon~\cite{Falkowski}.  In the second column, the first contribution to the uncertainty is due to neglected higher-order terms, and the second contribution is due to the combined effect of varying the factorization and renormalization scales.  The assumed axigluon mass is listed, and its assumed width is \SI{25}{\percent} of the mass in both cases.  The selection requirements imposed on all three calculations match the event selection requirements employed in our analysis: pseudorapidity $\abs{\eta_{b, \bar{b}}} < 1.1$.}

\begin{tabular}{cccc}
\hline
\hline
\\[-1.5ex]
\mbb range & \multicolumn{3}{c}{\afbb [\si{\percent}]} \\[0.5ex]
[\si[per-mode=symbol]{\giga\eV\per\clight\squared}] & SM & \gevM{200} & \gevM{345} \\[0.5ex]
\hline
\\[-1.3ex]
\([150, 225]\)  & \(2.43 \pm 0.73 ^{+0.02}_{-0.01}\) &   \(-2.9^{+0.4}_{-0.9}\) &   \(-1.9^{+0.4}_{-0.9}\) \\[0.5ex]
\([225, 325]\)  & \(4.61 \pm 1.38^{+0.15}_{-0.13}\) &   \(20.4^{+0.7}_{-1.0}\) &   \(-9.9^{+0.6}_{-0.6}\) \\[0.5ex]
\([325, 1960]\) & \(8.70 \pm 2.61^{+0.61}_{-0.51}\) &   \(20.2^{+0.4}_{-0.6}\) &   \(16.4^{+0.4}_{-0.9}\) \\[0.5ex]
\hline
\hline
\end{tabular}

\label{tab:GMcalc}
\end{table}

The forward-backward asymmetry for fermion-antifermion production is defined as
\begin{equation}
\label{e:afbdef}
\afb = \frac{n_F-n_B}{n_F+n_B},
\end{equation}
where $n_F$ is the number of events where the fermion is forward of the antifermion in rapidity~\cite{geometry} ($\Delta y = y_b - y_{\bar{b}} > 0$) and $n_B$ the number where it is backward ($\Delta y < 0$).  This definition is invariant under boosts along the beam axis.  

At hadron colliders, \bbbar pairs are almost exclusively produced by the strong interaction of quarks and gluons (QCD).  The vast majority of \bbbar pairs are produced via gluon-gluon fusion, which yields no asymmetry due to the symmetric initial state.  In the \(q\bar{q} \to \bbbar\) process, a positive \afbb arises from higher-order QCD corrections involving either real or virtual gluons.  Since the valence-quark parton-density functions dominate over the gluon parton-density functions at large Bjorken \(x\), we enhance the contribution of the quark-antiquark initial state over the symmetric gluon-fusion background by requiring large values of \bbar mass, $\mbb > \gevM{150}$.

A number of theoretical predictions of the SM value for \afbb at high mass have been reported, using various techniques and kinematic requirements and yielding a range of predictions~\cite{KuhnRodrigo99,ManoharEWK,Grinstein,*Murphy}.  K\"{u}hn and Rodrigo~\cite{KuhnRodrigo99} computed that the asymmetry for \bbbar pairs with \(\sqrt{\hat{s}} \geq \SI{300}{\giga\eV}\) and production angle \(\abs{\cos \theta^*} < 0.9\) falls in the range \SIrange{4.3}{5.1}{\percent}.  Manohar and Trott~\cite{ManoharEWK} computed \afbb in several bins of \mbb.  They found \(\afbb = \SI{0.4}{\percent}\) inclusively and \(\afbb = \) \SIrange{7.8}{8.1}{\percent} for \(350 < \mbb < \gevM{650}\), depending on the choice of the factorization and renormalization scales.  Grinstein and Murphy~\cite{Grinstein,*Murphy} compute predictions using a variety of kinematic requirements, including an estimate using our kinematic requirements. The results of this calculation~\cite{Grinstein,*Murphy} are summarized in \tabref{tab:GMcalc}.

Models containing a low-mass axigluon that has the same couplings to all quark flavours also predict a forward-backward asymmetry in bottom-quark-pair production. This asymmetry arises from the interference between the gluon and axigluon in the same fashion as the \(Z/\gamma^*\) interference of the Drell-Yan process~\cite{CDFDrellYanAfb,*DrellYan}.  One key feature of this interference is that the $\afb$ changes sign at the mass pole of the heavy axial resonance~\cite{AguilarLightAxiG,*Schmaltz,*Falkowski,*Ipek}.  Observation of such a sign flip would provide significant indication of the underlying dynamics.  The predictions of two different representative axigluon models are given in \tabref{tab:GMcalc}~\cite{Falkowski}. A relatively light \gevM{200} axigluon has been previously studied in the context of the top-quark asymmetry~\cite{CDFLJAfbl}, and a heavier \gevM{345} axigluon is of interest because its mass is just below the \ttbar threshold.  Both of these axigluons are assumed to have a width equal to \SI{25}{\percent} of their mass.

The predictions of both the SM and the models of non-SM physics under consideration are estimated at the parton level, and do not explicitly include the effects of hadronization.

Although the LHCb experiment has measured a related quantity~\cite{LHCbBBAC}, \afb can only be measured in proton-antiproton collisions~\footnote{The forward-backward asymmetry must vanish in a symmetric proton-proton initial state.}.  At the Tevatron, the D0 experiment has measured the forward-backward asymmetry of $B^{\pm}$ meson production~\cite{D0BmesonAfb}.  The momenta of the $B^{\pm}$ mesons in that measurement constrain the masses of the \bbbar pairs to be smaller than the range probed in this measurement.

\section{Experimental Apparatus and Event Selection}
\label{sec:evsel}

The Fermilab Tevatron is a proton-antiproton collider with a center-of-mass energy \(\sqrt{s} = \SI{1.96}{\tera\eV}\).  The CDF II detector~\cite{Acosta:2004yw} is an azimuthally symmetric magnetic spectrometer with a large tracking volume inside a solenoid.  Outside the solenoid are sampling calorimeters.  The calorimeters are further surrounded by the steel flux return of the solenoid and muon detectors~\cite{Acosta:2004yw}.

The data used for this analysis were collected by the CDF II detector using three online event selections (triggers), which require at least one jet with transverse energy \(E_T > 50\), 70, or \gevE{100}, respectively~\cite{geometry}.  To control the trigger rate, the two lower-threshold triggers only accept one event out of every 100 or 8 events that satisfy their requirements, respectively, while the highest-threshold trigger accepts all events meeting its requirements.  After data-quality requirements and trigger acceptance rates, the integrated luminosities of the samples are \SI{95}{\per\pico\barn}, \SI{1.2}{\per\femto\barn}, and \SI{9.5}{\per\femto\barn}, respectively.

The offline event selection requires at least two jets~\cite{JetConeSize,JetEnergyScale} with \(E_T > \gevE{20}\) and pseudorapidity \(\abs{\eta} < 1.1\).  Of the jets that pass these requirements, exactly two must contain a secondary vertex (separated transversely from the primary \ppbar interaction vertex) consistent with the decay of a $b$ quark ($b$-tagged jets), identified using a secondary-vertex identification algorithm~\cite{secvtx}.  The observed invariant mass \mbb of the two $b$-tagged jets is required to be at least \gevM{150}.

In order to maximize the statistical significance of the analysis, we separate the data into several subsamples depending on \mbb (three subsamples, see \secref{sec:meth}), on the estimated charges of the partons that produced the jets (four subsamples, see \secref{sec:jetQ}), and on the quality of the $b$ tags (three subsamples, see \secref{sec:purity}), yielding a total of 36 subsamples.
\section{Methodology}
\label{sec:meth}

We discriminate jets originated from $b$ quarks ($b$ jets) from jets originated from $\bar{b}$ quarks ($\bar{b}$ jets) using the momentum-weighted average of the charges of the particles associated with each jet (jet charge),
\begin{equation}
Q_\text{jet} = \frac{\sum_i q_i (\vec{p}_i \cdot \vec{p}_\text{jet})^{0.5}}{\sum_i (\vec{p}_i \cdot \vec{p}_\text{jet})^{0.5}},
\label{Qjet}
\end{equation}
where the sum $i$ is over all tracks in the jet, $q_i$ is the charge of the corresponding particle, and $\vec{p}_i$ and $\vec{p}_\text{jet}$ are the momentum vectors of the particle and of the jet, respectively.  The exponent $0.5$ was chosen to maximize the power of the jet charge in separating $b$ jets from $\bar{b}$ jets~\cite{TopCharge}.

The asymmetry depends on the mass of the \bbbar pair, both in the SM and in models with an axigluon.  To study the behavior of \afb as a function of \mbb, we divide the sample into several ranges (bins) of \mbb.  The choice of \mbb bins is motivated by the trigger.  Each jet-energy-trigger threshold efficiently selects events only over a limited range of \mbb.  Events with \(150 < \mbb < \gevM{225}\) are selected with a jet transverse-energy threshold of \SI{50}{\giga\eV}, and the thresholds of 70 and \SI{100}{\giga\eV} are used to select events with \mbb in the ranges \(225 < \mbb < \gevM{325}\) and \(\gevM{325} < \mbb\), respectively.

The measurement of \afbb must account for non-\bbbar backgrounds, detector effects, and the dilution of the asymmetry due to misidentification of $b$ jets as $\bar{b}$ jets and vice versa.  We therefore define a detector-level $\afb$,
\begin{equation}
\afb^\text{det} = \frac{1}{2P-1}\frac{(N_F - N^{\text{bkgd}}_F) - (N_B - N^{\text{bkgd}}_B)}{N_F - N^{\text{bkgd}}_F + N_B - N^{\text{bkgd}}_B},
\label{eq:masterequation}
\end{equation}
where $N_F$ and $N_B$ are the observed numbers of events, $N^{\text{bkgd}}_F$ and $N^{\text{bkgd}}_B$ are the estimated numbers of background events, and $P$ is the probability to make the charge assignment correctly.  To account for additional effects, such as the finite resolution and acceptance of the CDF II detector, we employ a Bayesian technique to measure the \afb at particle level.  ``Particle level'' refers to quantities reconstructed from final-state, color-confined particles with lifetimes greater than \SI{10}{\pico\second}~\cite{hadronjets}.

The background levels are determined with a data-driven technique based on the $b$-tagged vertex mass as described in \secref{sec:purity}. The calibration of the charge misidentification and the calculation of $P$ is presented in \secref{sec:jetQ}, and the background asymmetries are discussed in \secref{sec:background}. Finally, we use a Markov-chain Monte Carlo calculation to derive particle-level results by identifying the maxima of the marginalized posterior probability densities and by constructing associated credible intervals. The correction to the particle level, including the effects of mismeasurement and acceptance, are described in \secref{sec:bayes}. 

\subsection{Identification of the \texorpdfstring{\boldmath{$b$}}{b} jet using jet charge}
\label{sec:jetQ}

\begin{figure}
\includegraphics[width=0.45\textwidth]{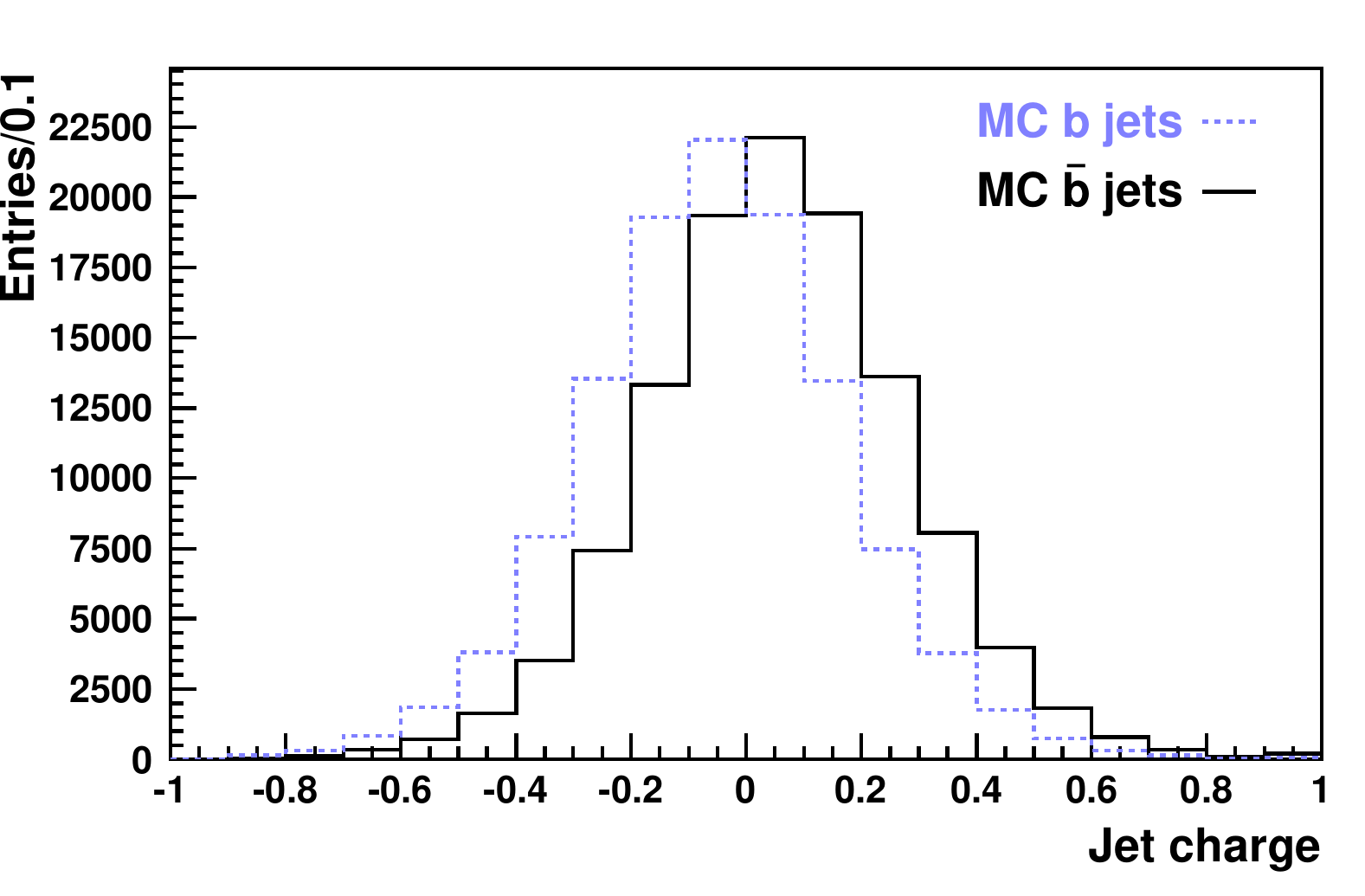}
\caption{Distributions of the charge of $b$ and $\bar{b}$ jets in Monte Carlo simulation of two-jet events.}
\label{f:jetq}
\end{figure}

The forward or backward assignment is performed using the momentum-weighted track charge, or ``jet charge'' (see \eqref{Qjet}), for each of the two $b$-tagged jets.  Distributions of the jet charge ($Q_\text{jet}$) for $b$ jets and $\bar{b}$ jets are shown in \figref{f:jetq}.  We use the difference of the two jet charges \(\Delta Q = Q_1 - Q_2\) to make the assignment: if $\Delta Q$ is negative, the jet with charge $Q_1$ is considered to be the $b$ jet (because the $b$ quark is negatively charged), and if $\Delta Q$ is positive, the jet with charge $Q_2$ is considered to be the $b$ jet.

The performance of the jet-charge algorithm is calibrated with data, by dividing the distribution of $Q_\text{jet}$ into bins.  The bin edges are \numlist[list-final-separator={, and }]{-0.25;0;0.25}, for a total of four bins (the lowest- and highest-charge bins are open).  We arbitrarily assign $Q_\text{jet}$ values of \numlist[list-final-separator={, and }]{-0.5;-0.25;0.25;0.5} to the jets falling into each respective bin.  This translates into five bins of $\abs{\Delta Q}$ with values \numlist[list-final-separator={, and }]{0;0.25;0.5;0.75;1.0}.  The bin with $\abs{\Delta Q} = 0$ is not informative because it gives no indication of how to assign the $b$ jet, so only four bins of $\abs{\Delta Q}$ are used.

\subsection{Charge-identification probability}

The probability $P$ of correctly assigning the $b$ jet and $\bar{b}$ jet, introduced in \eqref{eq:masterequation}, is estimated from the data.  We evaluate this in each bin of $\abs{\Delta Q}$, so that
\begin{eqnarray}
P_{0.25} & = & \frac{p_{0.5}(1-p_{0.25})}{p_{0.5}(1-p_{0.25})+(1-p_{0.5})p_{0.25}}, \nonumber\\
P_{0.5} & = & \frac{p_{0.25}^2}{p_{0.25}^2+(1-p_{0.25})^2}, \nonumber\\
P_{0.75} & = & \frac{p_{0.25}p_{0.5}}{p_{0.25}p_{0.5}+(1-p_{0.25})(1-p_{0.5})},\quad\text{and} \nonumber\\
P_{1.0} & = & \frac{p_{0.5}^2}{p_{0.5}^2+(1-p_{0.5})^2},
\label{chargeprobdef}
\end{eqnarray}
where the subscript of $P$ indicates \(\abs{\Delta Q}\), and the various $p_{\abs{Q_\text{jet}}}$ are the probabilities that a jet with a binned charge of $\abs{Q_\text{jet}}$ is correctly identified as $b$ or $\bar{b}$ by the sign of $Q_\text{jet}$.  These expressions are derived by exhaustively considering all the possibilities.  For example, \(\abs{\Delta Q} = 1.0\) implies that one jet has \(Q_\text{jet} = +0.5\) and the other jet has \(Q_\text{jet} = -0.5\).  If the \(b\) jet has charge \num{-0.5}, then the use of $\Delta Q$ allows for a correct assignment of the \(b\) and \(\bar{b}\) jets.  This case occurs with probability $p_{0.5}^2$, because both jets (with \(\abs{Q_\text{jet}} = 0.5\)) have a jet charge whose sign matches the sign of the charge of the originating quark.  The other possibility is that we misidentify both jets, which occurs with probability \((1-p_{0.5})^2\).  The denominator in \eqref{chargeprobdef} is the sum of these two probabilities, and the numerator only contains the correct case.

In order to measure $p_{\abs{Q_\text{jet}}}$, we measure the number of opposite-charge events, $N_\text{OC}$, and number of same-charge events, $N_\text{SC}$, in the data in each \mbb subsample. These numbers are corrected for the presence of background using the calibrated $b$-fractions (see \secref{sec:purity}) and the background model described in \secref{sec:background}.  We assume that the remaining sample is composed of \bbbar with no contamination from $bb$ or $\bar{b}\bar{b}$, and compute the opposite-charge fraction $F_\text{OC} = N_\text{OC}/(N_\text{OC}+N_\text{SC})$.

In events in which both jets have $\abs{Q_\text{jet}} = 0.25$, the opposite-sign fraction is expressed as
\begin{equation}
F^{0.25-0.25}_\text{OC} = p_{0.25}^2 + (1-p_{0.25})^2,
\end{equation}
where the term $p^2_{0.25}$ arises from events in which both $b$ jets have the correct sign, and the term $(1-p_{0.25})^2$ arises from events in which both $b$ jets have the wrong sign.  Similarly, we have
\begin{eqnarray}
F^{0.25-0.5}_\text{OC} & = & p_{0.25}p_{0.5}+(1-p_{0.25})(1-p_{0.5}),\quad\text{and} \\
F^{0.5-0.5}_\text{OC} & = & p_{0.5}^2 + (1-p_{0.5})^2,
\end{eqnarray}
for events containing one jet with $\abs{Q_\text{jet}} = 0.25$ and one jet with $\abs{Q_\text{jet}} = 0.5$, and for events in which both jets have $\abs{Q_\text{jet}} = 0.5$, respectively.  We measure each $F_\text{OC}$ value in the data and solve for each $p_{\abs{Q_\text{jet}}}$, and then for each of the four $P_{\abs{\Delta Q}}$.

\subsection{Sample purity}
\label{sec:purity}

\Eqref{eq:masterequation} also requires knowledge of the rate of background events, $N^\text{bkgd}_F$ and $N^\text{bkgd}_B$.  We obtain the background yields by estimating the number of \bbbar events $N_{\bbbar}$ and subtracting it from the number of events observed.

We divide the data into subsamples with varying $b$-tag quality, which in turn provides subsamples of varying \bbbar purity and improves the statistical power of the measurement.  Since we apply $b$-tags to jets based on the presence of a secondary vertex, the quality of the $b$-tag is based on the confidence level of the identification of the secondary vertex.  Specifically, the significance is based on the distance $L_\text{2D}$, in the plane perpendicular to the beam, between the primary and secondary vertices, projected onto the jet momentum~\cite{secvtx}.  Jets with a significance ($\abs{L_\text{2D}} / \sigma(L_\text{2D})$) greater than 20 are high significance or ``H'' tags, and jets with a lower significance are ``L'' tags.  This results in the following independent subsamples ranked in order of increasing \bbbar purity: LL (both jets are L tagged), LH (one L and one H tag), and HH (both H tags).

We estimate the number of events in which both $b$-tagged jets are genuine $b$ jets by counting events in which one or both of the $b$-tagged jets have a negative $L_\text{2D}$.  These ``negative'' tags are predominantly false tags from light-flavor jets and are a consequence of the finite position resolution of the tracking system.  We expect the rate of false tags from this source to be equal for positive and negative tags.  There are additional false positive tag contributions from hyperon decays and from interactions between high-momentum particles and the detector material.  This results in an excess of positive over negative false tags.  We exploit this relationship between the number of positively- and negatively-tagged light-flavor jets and compute the number of true \bbbar events using
\begin{equation}
\label{e:nonbb}
N_{\bbbar} = \frac{1}{\xi}(N_{++} - \lambda N_{+-} + \lambda^2 N_{--}),
\end{equation}
where $N_{++}$ is the number of observed positive double-tag events, $N_{+-}$ is the number of events with one of the tags negative, $N_{--}$ is the number with both tags negative, $\lambda$ is the ratio of positive to negative false tags, and \(\xi = 1 - 2\lambda r + \lambda^2 r^2\) is a factor that accounts for the presence of negatively-tagged $b$ jets. Here, $r = \epsilon_b^-/\epsilon_b^+$ is the ratio of the negative to positive tag efficiencies for $b$ jets, estimated using Monte Carlo simulation and data.  Finally, $N_{\bbbar}$ is corrected by a scale factor derived from MC to account for a bias resulting from the presence of charm jets~\cite{CDFSusyHiggs}.

\begin{figure}[tbhp]
\includegraphics[width=.45\textwidth]{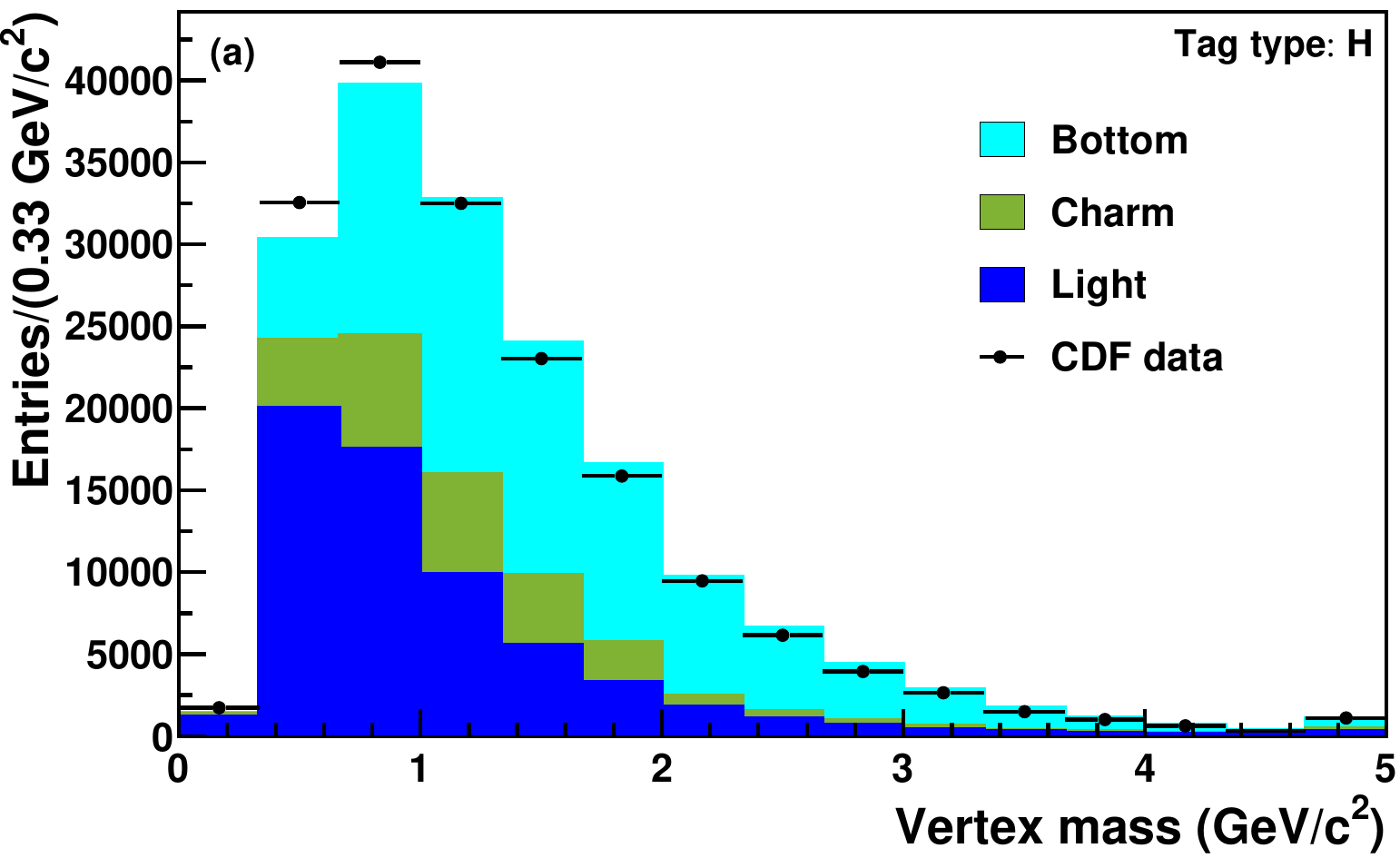}
\includegraphics[width=.45\textwidth]{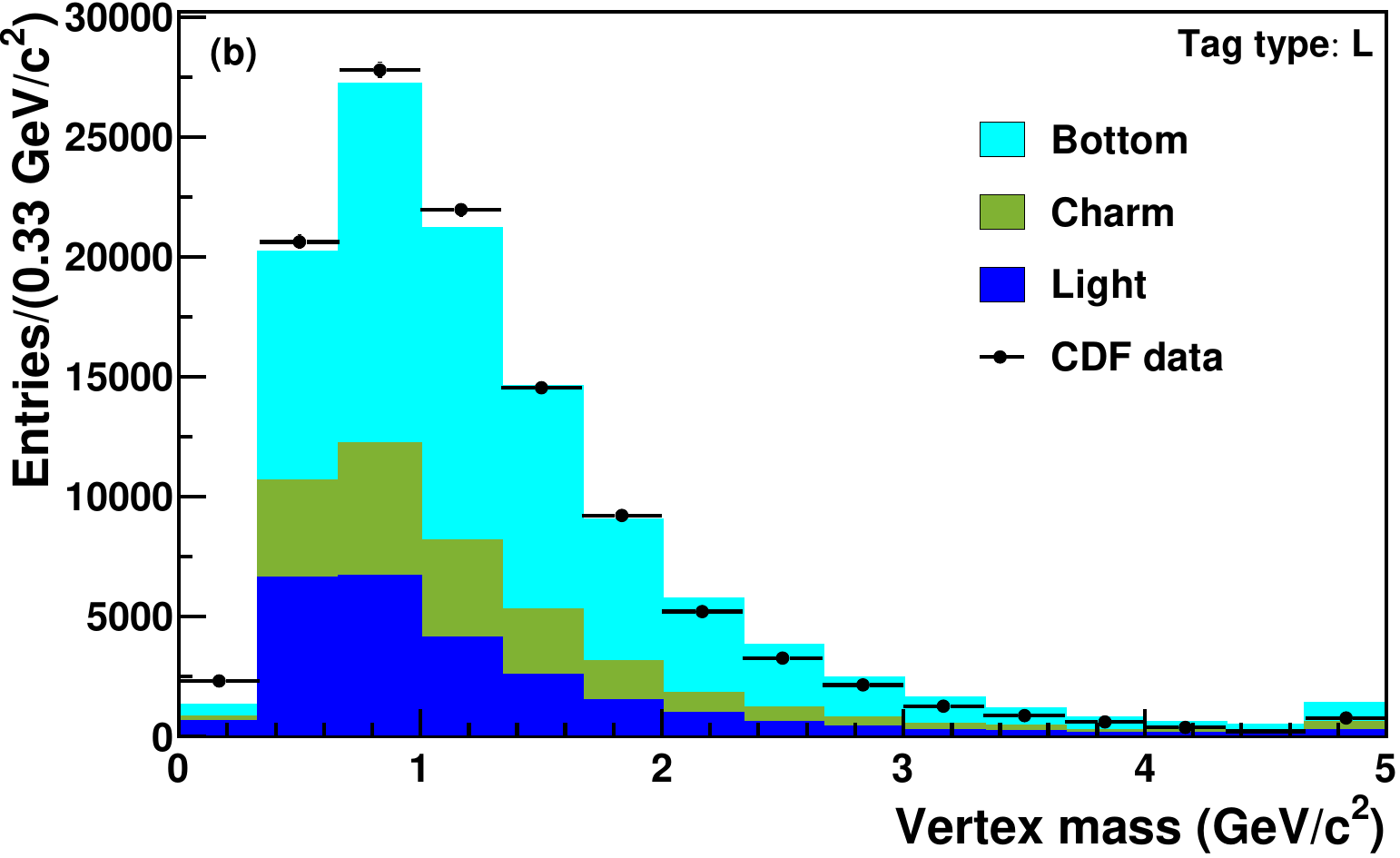}
\caption{Distribution of net $m_\text{VTX}$, for H (a) and L (b) tagged data, with fit overlaid.  The negative tag rates are subtracted from the positive tag rates.  The histograms are stacked on top of one another.}
\label{f:netfit}
\end{figure}

We measure the ratio of positive to negative light-flavor tag rates $\lambda$ following Ref.~\cite{CDFSusyHiggs}.  Since the MC indicates no strong dependence of $\lambda$ on jet $E_T$, we use the same value for all three dijet mass bins.  We study the distribution of the invariant mass of all charged particles associated with the secondary vertex, $m_\text{VTX}$~\cite{secvtx}, for both negative and positive tags.  We subtract the negative tag distribution from the positive tag distribution, yielding the ``net'' distribution.  The relative rates of MC-derived templates for $b$, $c$, and light-flavor jets, left as free parameters, are fit in the net $m_\text{VTX}$ distribution observed in data (see \figref{f:netfit}).

These net flavor fractions allow for the normalization of the complete positive- and negative-tagged distributions of $m_\text{VTX}$ for all jet flavors.  We find that the negative-tagged component of the MC templates must be scaled by 1.5 (1.3) to match the total number of low (high) significance negative tags observed in data.  This scaling is propagated into the positive-tagged region of the templates as this population of ``fake'' tags is considered to be positive-negative symmetric, leaving the original ``net'' distributions unchanged~\cite{secvtx}.  From these corrected templates, we then obtain $\lambda$ for low- and high-significance $b$-tagged jets.  We also correct the values of $r$ used in the three dijet mass bins for the higher negative tag rates observed in the data.

\subsection{Background model}
\label{sec:background}

\begin{figure}
\includegraphics[width=0.45\textwidth]{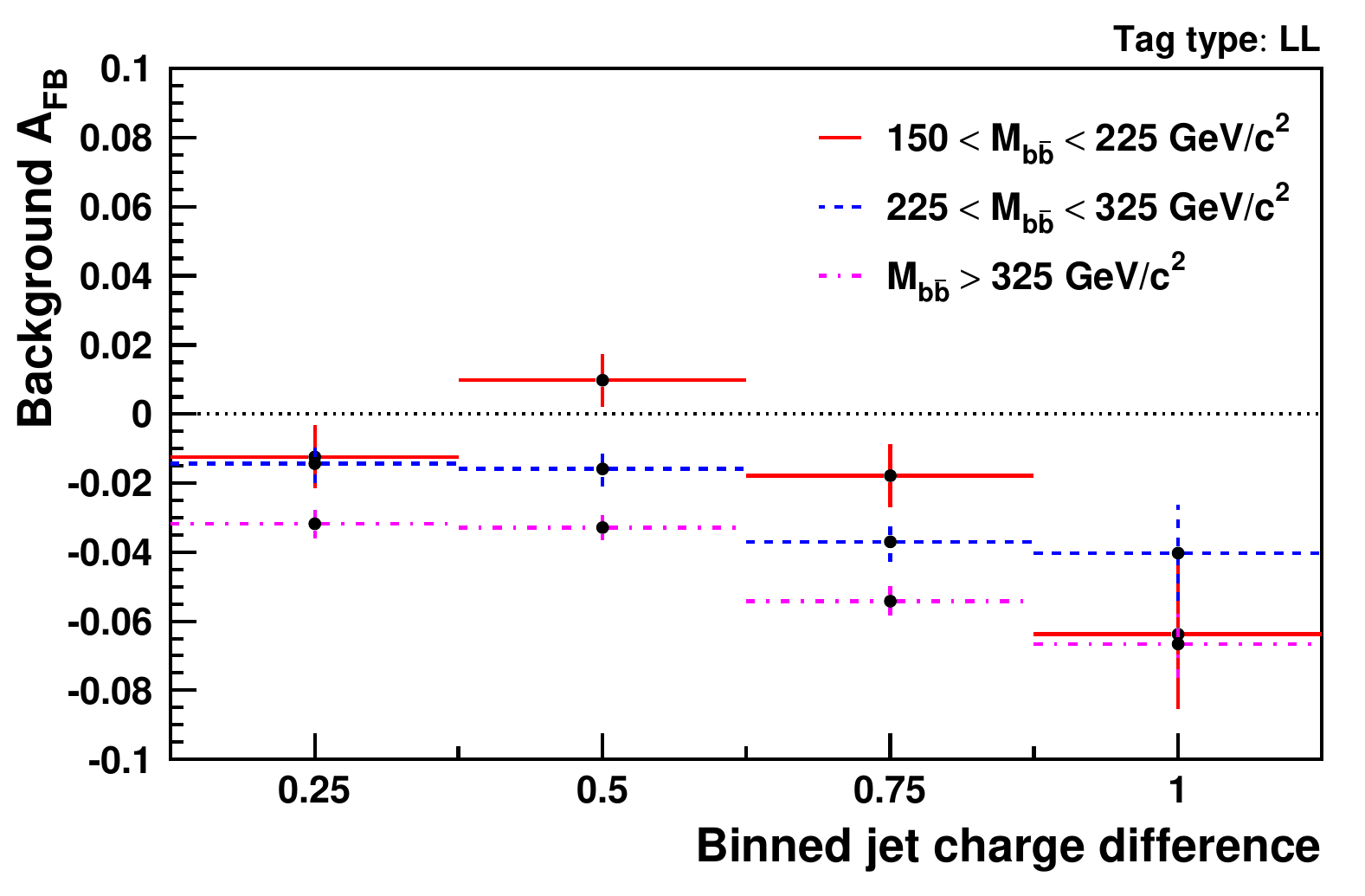}
\caption{Estimated asymmetry of the backgrounds in each \mbb bin with tag type LL, as a function of $\abs{\Delta Q}$.}
\label{f:basym}
\end{figure}

The non-\bbbar background includes a wide variety of physics processes, such as $b$+mistag, $u\bar{u}$, $d\bar{d}$, gluon jets, etc.  Of primary concern is the scattering of the valence quarks, which proceeds through the $t$-channel exchange of a gluon and exhibits a forward-backward asymmetry due to the forward Rutherford peak.  Rather than attempting to tune the simulation to reproduce all of these processes in the proper amounts, we use data in a control region that is expected to be enhanced in background and depleted of \bbbar signal.  The control region is defined by loosening the requirements of the $b$-tagging algorithm (looser than the low-significance ``L'' tag), and requiring that at least one of the tagged jets is negatively tagged.  The sample of events satisfying these criteria is non-overlapping with the signal sample, and is composed almost entirely of background but is kinematically very similar to the signal sample.  The forward-backward asymmetry of this sample is assumed to approximate the forward-backward asymmetry of the background in the signal sample (see \figref{f:basym}).

The asymmetry of the background is negative, consistent with $u$--$\bar{u}$ scattering as the most important component.  The $u$ quark tends to follow the incoming proton direction in $t$-channel scattering which would seem to yield a positive $\afb$, but the opposite charge of the $u$ and $b$ quarks, combined with the definition of the asymmetry in terms of the jet charge, reverses the sign of the asymmetry.

Together with the background yields, the estimated background asymmetries are used to estimate the number of forward and backward background events, $N^\text{bkgd}_F$ and $N^\text{bkgd}_B$, in \eqref{eq:masterequation}.

\subsection{Jet energy mismeasurement}

\begin{figure}
\includegraphics{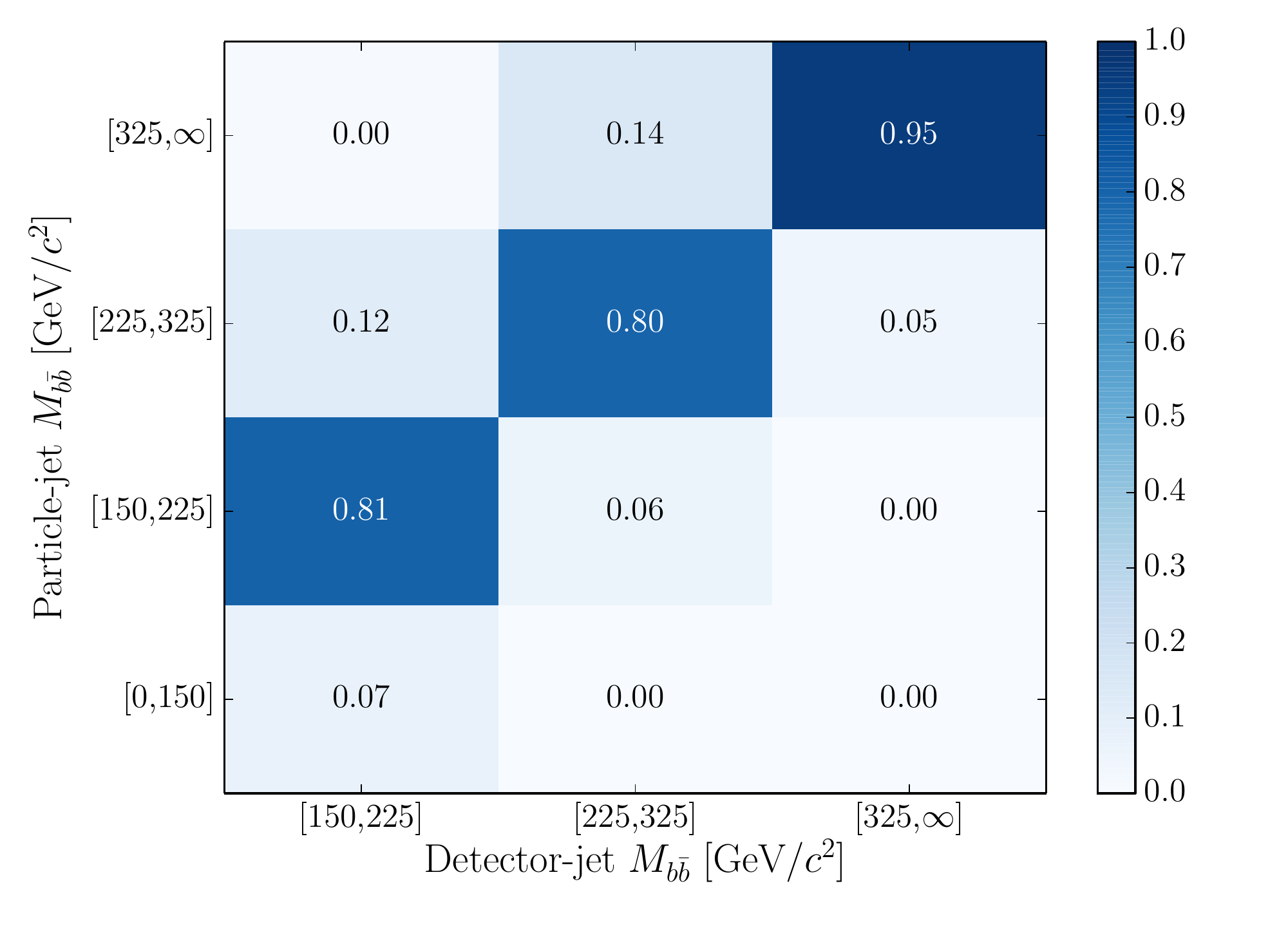}\caption{Smearing matrix relating measured dijet mass to particle-level dijet mass.  The matrix is evaluated from MC, and depends on $\abs{\Delta Q}$, but it is shown summed over bins of $\abs{\Delta Q}$.}
\label{Msmear}
\end{figure}

\begin{figure}
\includegraphics[width=0.45\textwidth]{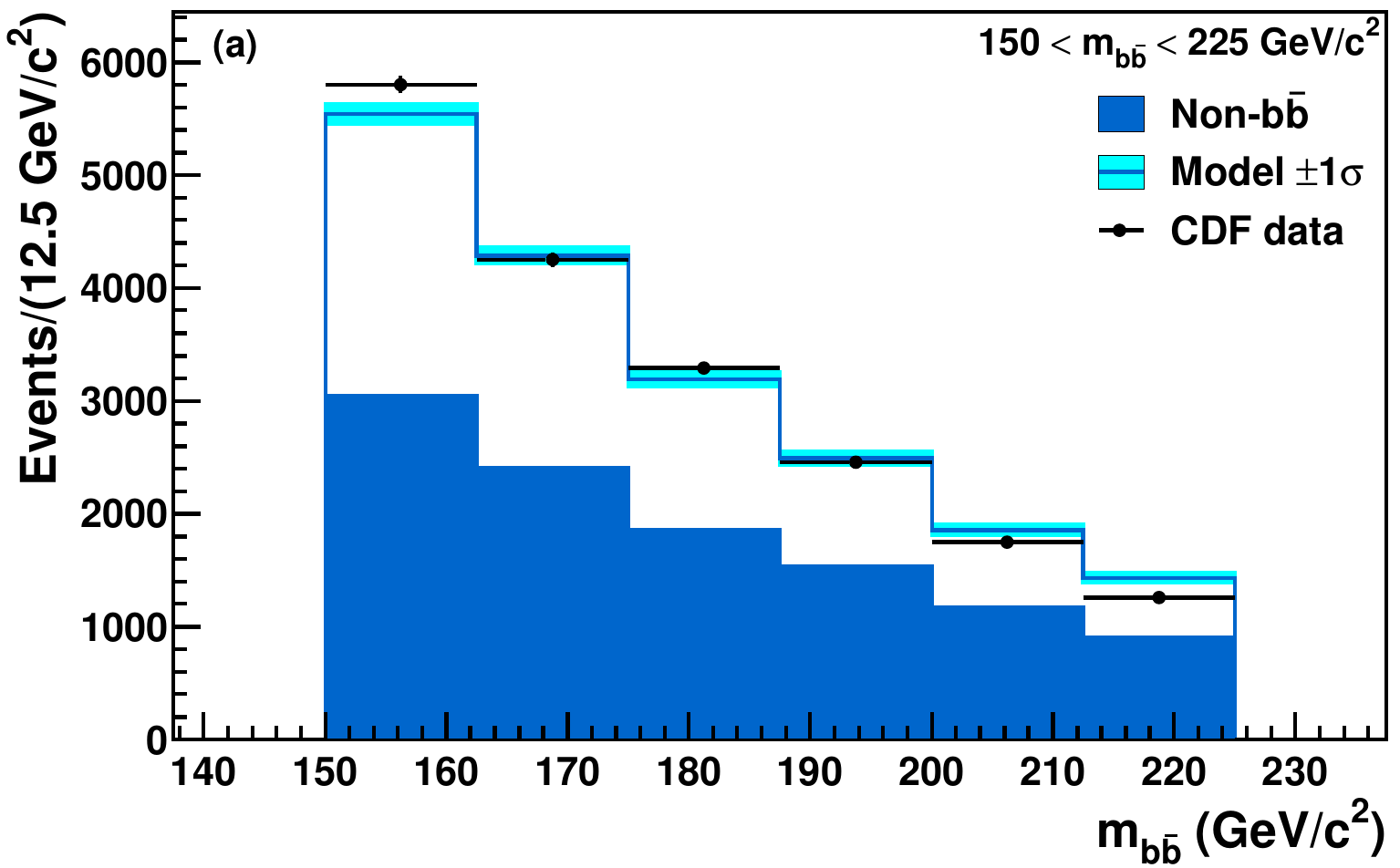}
\includegraphics[width=0.45\textwidth]{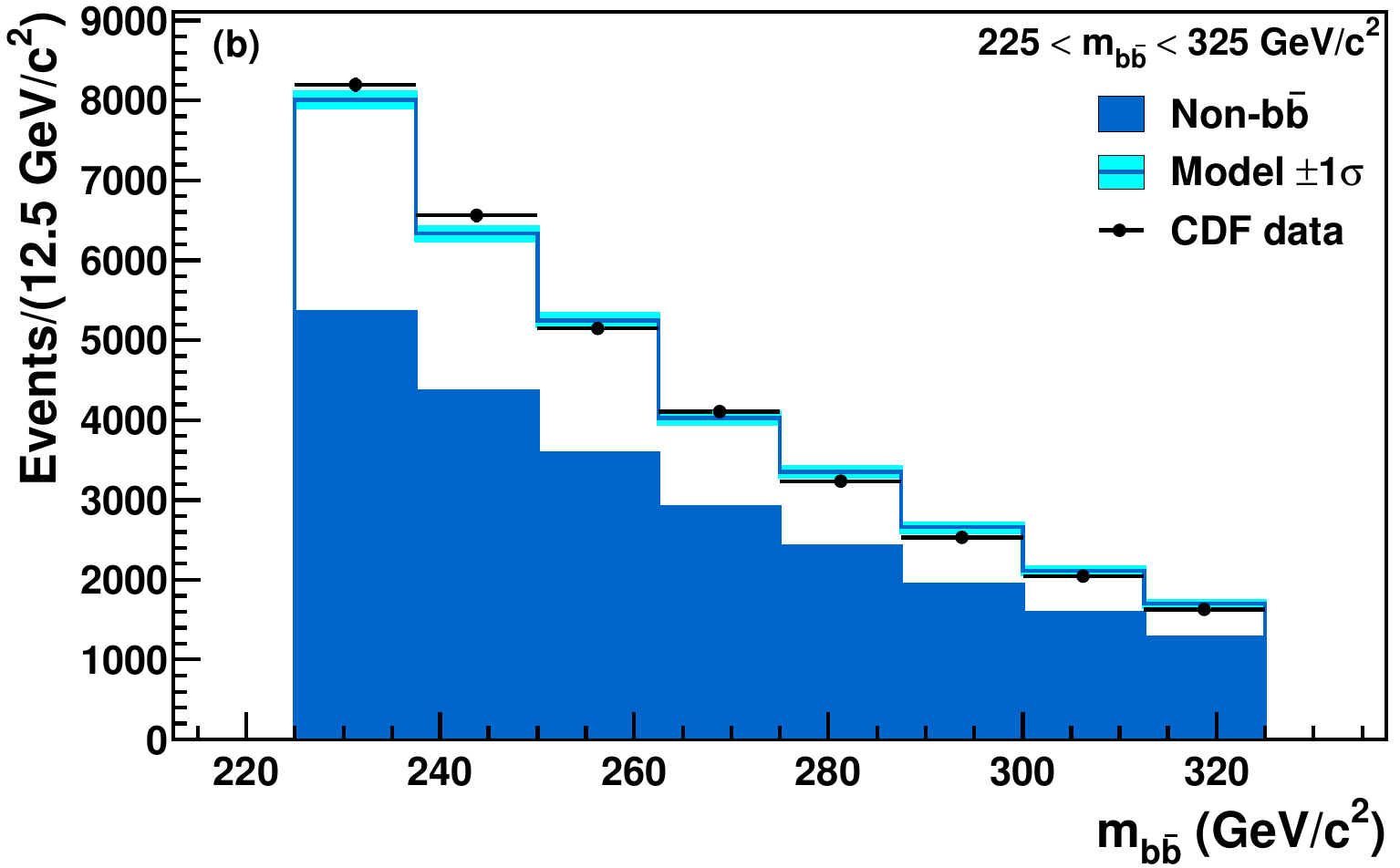}
\includegraphics[width=0.45\textwidth]{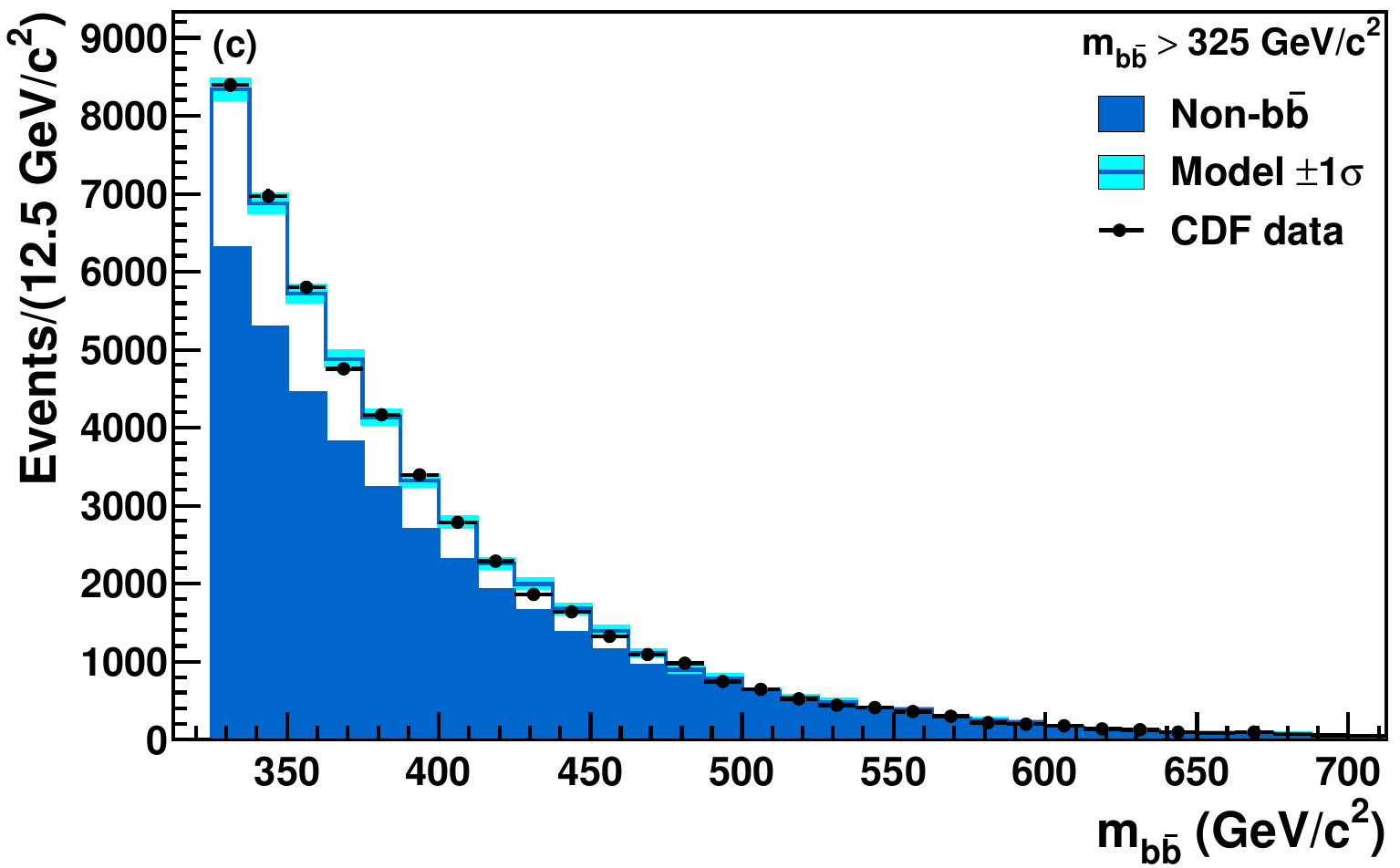}
\caption{Detector-level dijet mass spectra in each of the three \mbb subsamples, showing the spectrum resulting from the non-\bbbar component only as well as from the sum of the \bbbar and non-\bbbar components (``Model$\pm 1 \sigma$'').  The agreement between data and simulation is good, suggesting that the particle-level dijet mass in MC simulation and the smearing matrix are also correct within the uncertainties.  The non-\bbbar component is not included in the smearing matrix.}
\label{mjj}
\end{figure}

Although we correct the energy of each jet for known effects~\cite{JetEnergyScale}, the energy of a measured jet does not exactly match the energy of the corresponding particle-level jet~\cite{hadronjets}.  Mismeasurement of jet energies principally affects the measurement of the dijet mass.  Since we use wide mass bins, this effect is small.  We estimate the effect using MC samples of dijet events produced with the \textsc{pythia} event generator at leading order in the strong coupling constant, using the \textsc{cteq5l} parton distribution functions.  We select events that have a pair of bottom quarks, whether produced directly or in the parton shower, to produce the signal model.

This signal model allows the determination of a matrix relating the measured dijet mass to the particle-level dijet mass, as a function of $\abs{\Delta Q}$.  The matrix, summed over $\abs{\Delta Q}$, is shown in \figref{Msmear}.  In 80 to \SI{95}{\percent} of events, the dijet mass is reconstructed in the same bin.  The relative uncertainty on each element of the matrix ranges from a few percent on the diagonal up to very large values far off the diagonal.  The fraction of events in which the detector-level dijet mass migrates to a different bin than that of the particle-level dijet mass necessitates a correction, and increases the uncertainty in the measurement of the particle-level \afb as a function of dijet mass (see \secref{sec:bayes}).  The correction relies on the MC description of the particle-level dijet mass to be approximately correct.  \Figref{mjj} shows that the detector-level dijet mass is correctly described by the MC, which supports the reliability of the smearing matrix.

\subsection{Selection effects}
\label{sec:acc}

We also estimate the effect of the potentially asymmetric acceptance of the detector and analysis selection.  We apply the analysis event selection to the simulated events to estimate the fractions of forward ($\epsilon_F$) and backward ($\epsilon_B$) events that pass the analysis selection.  We perform this estimation in each bin of particle-level dijet mass.  Only the ratio of the forward to the backward acceptance affects the analysis, so we compute the ratio $R = \epsilon_F / \epsilon_B$.  We also estimate the uncertainty on this ratio by varying the renormalization and factorization scales used in the signal model by a factor of two and the jet-energy scale within its uncertainty.

\subsection{Extraction of the particle-level asymmetry}
\label{sec:bayes}

\begin{table}


\caption{Numbers of events observed in the various subsamples.  }

\begin{tabular}{cccSS}
\hline
\hline
\mbb range & \(\Delta Q\) & tag type & {\(N(\Delta y > 0)\)} & {\(N(\Delta y < 0)\)} \\
\hline
\( [150, 225]\) & 0.25 & LL &  483 &  472 \\
                &      & LH &  774 &  841 \\
                &      & HH &  465 &  469 \\
                &  0.5 & LL &  734 &  699 \\
                &      & LH & 1223 & 1240 \\
                &      & HH &  682 &  691 \\
                & 0.75 & LL &  460 &  474 \\
                &      & LH &  854 &  882 \\
                &      & HH &  483 &  519 \\
                &  1.0 & LL &   97 &   81 \\
                &      & LH &  130 &  141 \\
                &      & HH &  101 &  106 \\
\( [225, 325]\) & 0.25 & LL & 1014 &  984 \\
                &      & LH & 1419 & 1520 \\
                &      & HH &  686 &  735 \\
                &  0.5 & LL & 1465 & 1499 \\
                &      & LH & 2197 & 2230 \\
                &      & HH &  951 &  992 \\
                & 0.75 & LL &  915 &  979 \\
                &      & LH & 1455 & 1558 \\
                &      & HH &  712 &  735 \\
                &  1.0 & LL &  140 &  160 \\
                &      & LH &  293 &  256 \\
                &      & HH &  149 &  140 \\
\([325, 1960]\) & 0.25 & LL & 1565 & 1636 \\
                &      & LH & 2214 & 2401 \\
                &      & HH &  965 &  937 \\
                &  0.5 & LL & 2228 & 2379 \\
                &      & LH & 3196 & 3254 \\
                &      & HH & 1298 & 1318 \\
                & 0.75 & LL & 1585 & 1680 \\
                &      & LH & 2286 & 2512 \\
                &      & HH &  958 &  963 \\
                &  1.0 & LL &  253 &  281 \\
                &      & LH &  396 &  439 \\
                &      & HH &  189 &  191 \\
\hline
\hline
\end{tabular}

\label{tab:yield}
\end{table}

To infer the asymmetry at the particle level, we construct a Bayesian model to describe the data.  This model combines all effects discussed in the preceding sections, propagates the uncertainties, and allows the data to constrain the uncertainties when possible.  The parameters of the model and their assumed prior probability distributions are as follows:

\begin{enumerate}
\item $f_{MQT}$ is the \bbbar fraction in each bin of detector-level dijet mass $M$, charge difference $Q$, and tag-quality $T$.  The prior probability distribution for $f$ is a normal distribution centered at the calibrated value, with a width equal to the residual uncertainty from the calibration.
\item $P_{MQ}$ is the correct charge probability described in \eqref{chargeprobdef}.  The priors for $p_{0.5}$ and $p_{0.25}$ are normal distributions with mean and uncertainty taken from the calibration.
\item $F^\text{bkgd}_{MQT}$ and $B^\text{bkgd}_{MQT}$ are the rates of forward and backward events in the background-dominated sideband.  From these, we calculate the background asymmetry, which is assumed to be consistent with the asymmetry of the background in the signal region.  The prior for each of these is the gamma distribution.
\item $J$ is the shift in the jet energy scale.  We coherently shift all of the jet energies in every MC event by $J$ times the jet-energy uncertainty~\cite{JetEnergyScale}.  The prior is a normal distribution with a mean of zero and standard deviation of one.
\item $S_{M'MQ}(J)$ is a matrix describing the contribution of events with particle-level dijet mass $M'$ to the various bins of measured dijet mass and charge difference.  The matrix is a function of $J$.  The prior is taken from the rate and uncertainty in simulation, and the matrix is normalized so that $\sum_{M'} S_{M'MQ}(J) = 1$.
\item $\sigma_{MQT}$ is the rate of events in each bin of detector-level \bbbar mass, charge difference, and tag quality.  This parameter has a uniform prior over the nonnegative range.  This parameter is necessary because the simulation does not accurately predict the overall event rate as a function of mass, charge difference, and tag quality.
\item $R_{M'}$ is the ratio of the forward to the backward acceptance.  The prior is a normal distribution with mean and width taken from the calibration described in \secref{sec:acc}.
\item $A_{M'}$ is the \bbbar asymmetry in bins of particle-level mass.  This is the parameter we wish to measure.  We use a uniform prior from $[-1, 1]$.
\item $A^\text{acc}_{M'}$ is the \bbbar asymmetry after acceptance and selection effects.  It is a function of $A_{M'}$ and $R_{M'}$, \[A^\text{acc}_{M'} = \frac{R_{M'}(1+A_{M'}) - (1-A_{M'})}{R_{M'}(1+A_{M'}) + (1-A_{M'})}.\]
\end{enumerate}

We compute the rate $\theta$ of forward and backward events expected in data, and compare this rate to that observed in data (see \tabref{tab:yield}) via a Poisson likelihood, given an observation of $k$ events,
\([L(\theta | k) = \theta^k e^{-\theta} / k!\).
The rates are
\begin{align}
\theta^\text{Forward}_{MQT} &= \Biggl[f_{MQT} \sum_{M'} \frac{1 + A^\text{acc}_{M'} (2 P_{MQ} - 1)}{2} S_{M'MQ}(J) \nonumber \\
&+ (1-f_{MQT}) \frac{F^\text{bkgd}_{MQT}}{F^\text{bkgd}_{MQT} + B^\text{bkgd}_{MQT}}\Biggr] \times \sigma_{MQT},
\end{align}
and
\begin{align}
\theta^\text{Backward}_{MQT} &= \Biggl[f_{MQT} \sum_{M'} \frac{1 - A^\text{acc}_{M'} (2 P_{MQ} - 1)}{2} S_{M'MQ}(J) \nonumber \\
&+ (1-f_{MQT}) \frac{B^\text{bkgd}_{MQT}}{F^\text{bkgd}_{MQT} + B^\text{bkgd}_{MQT}}\Biggr] \times \sigma_{MQT}.
\end{align}
The prior probability densities described above, together with this likelihood, fully specify the posterior probability density for the parameters.  To estimate this posterior density, we employ Markov-chain Monte Carlo sampling~\cite{pymc}.  This technique provides us with samples from the posterior probability distribution over the parameter space.  We marginalize the nuisance parameters and obtain the posterior density for $A_{M'}$, the asymmetry in each bin of particle-level \bbbar mass.  The marginal distributions are shown in \figref{fig:MarginalPosteriorAFB}, and the marginalized samples are provided in the supplemental material~\footnote{See supplemental material at [URL will be provided by publisher]}.

\section{Results and Conclusions}

\begin{figure}
\includegraphics{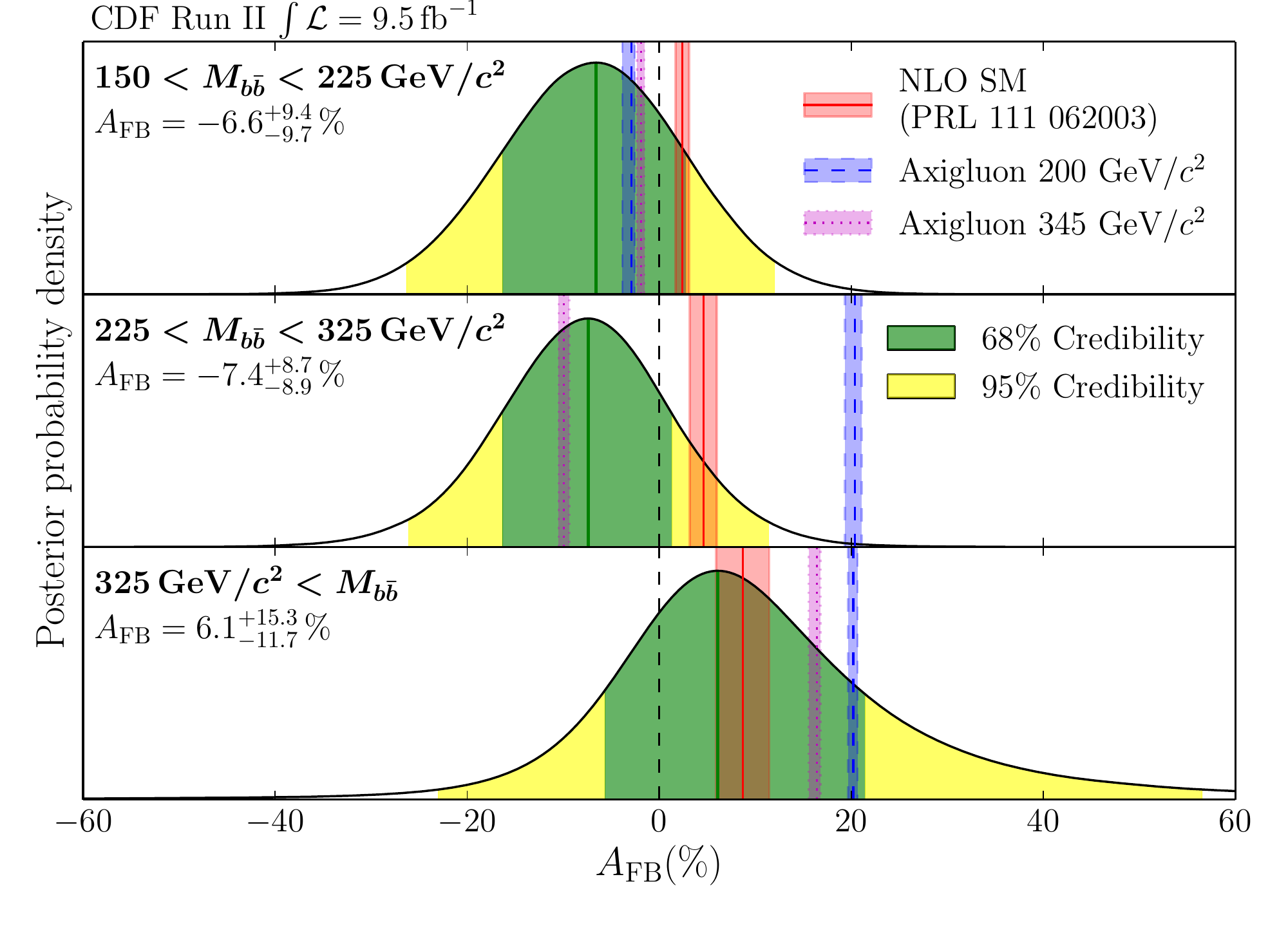}
\caption{Marginal posterior probability distribution of asymmetry in each bin of particle-level \bbbar mass.  The inner and outer bands represent the \SI{68}{\percent} and \SI{95}{\percent} credible intervals, respectively.}
\label{fig:MarginalPosteriorAFB}
\end{figure}

\begin{figure}
\includegraphics{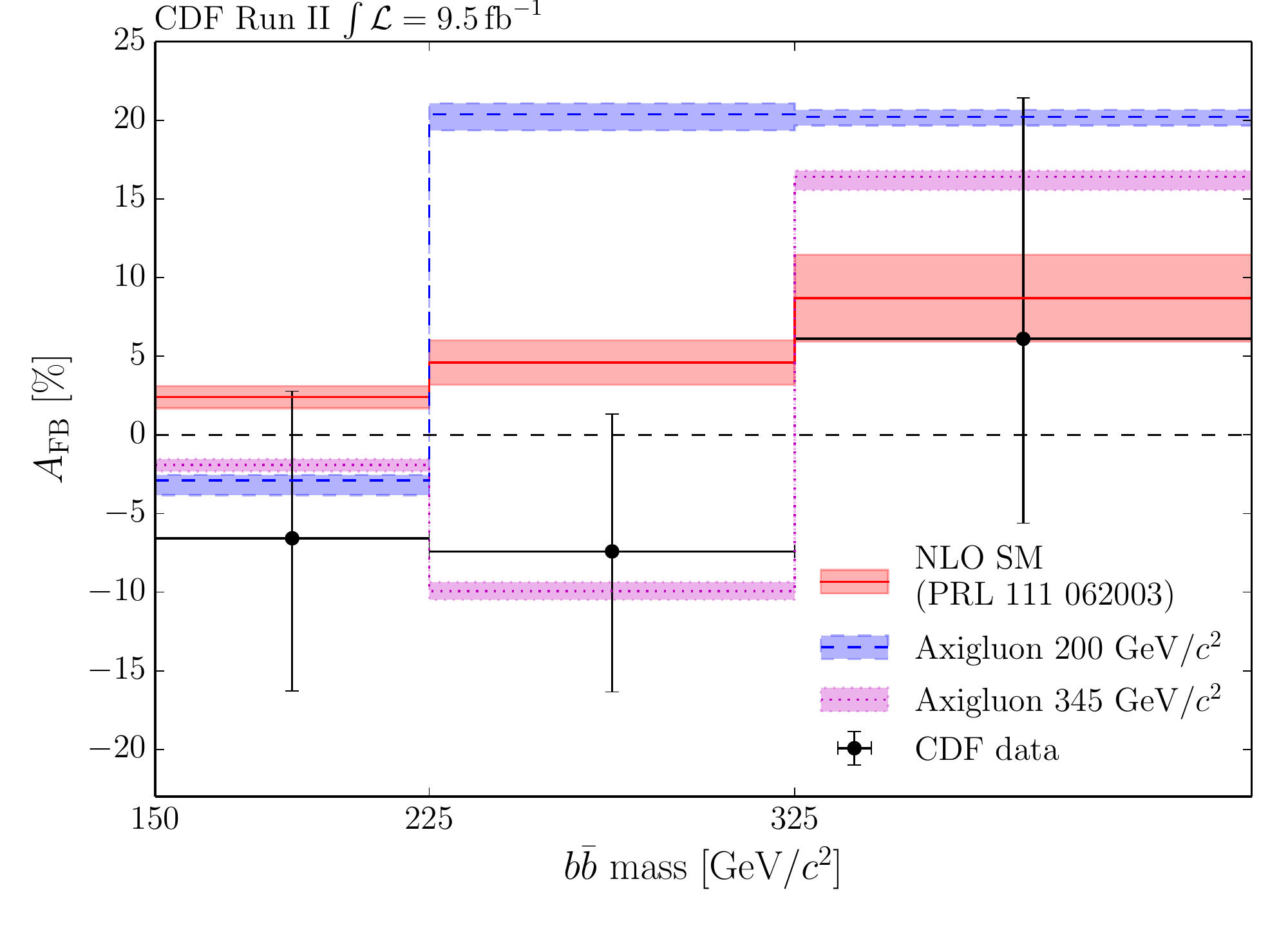}
\caption{Values of maximum \textit{a posteriori} signal asymmetry as a function of \bbbar mass. The error bars represent the \SI{68}{\percent} credible intervals.}
\label{fig:AFBvsMass}
\end{figure}

To characterize the posterior and describe the measurement, we find the highest probability-density credible intervals at \SI{68}{\percent} and \SI{95}{\percent} credibility for $A_{M'}$ in each particle-level mass bin.  The posterior densities, along with the intervals describing them, are shown in \figref{fig:MarginalPosteriorAFB}.  The red vertical bands (with solid lines) represent the theoretical predictions from the SM~\cite{Grinstein,*Murphy}, while the blue and magenta bands (with dashed and dotted lines, respectively) represent the predictions from two axigluon models (see \tabref{tab:GMcalc}).  Although the measurement is performed at the particle level, the predictions are at the parton level and do not include the effects of hadronization.  Because hadronization effects are not expected to be large, we do not hesitate to interpret the results.
  
The measured asymmetries, summarized in~\figref{fig:AFBvsMass}, are \(-6.6^{+9.4}_{-9.7}\)\si{\percent}, \(-7.4^{+8.7}_{-8.9}\)\si{\percent}, and \(6.1^{+15.3}_{-11.7}\)\si{\percent} in the low, middle, and high \mbb bins, respectively.  These results, which account for the effects of backgrounds, charge misidentification, detector resolution, and nonuniform detector acceptance, are consistent with zero and with the standard model prediction~\cite{Grinstein,*Murphy} in each bin.  Only \SI{0.24}{\percent} of the posterior probability density in the middle mass bin has an \afb larger than predicted by the lighter axigluon model~\cite{Falkowski} with a mass of \gevM{200}.  Accounting for the look-elsewhere effect following Ref.~\cite{SidakCorr}, this is sufficient to exclude the lighter axigluon at more than \SI{95}{\percent}.  The measurement is unable to exclude the heavier axigluon with a mass of \gevM{345}.  This measurement reduces the allowed parameter space for light axigluon models used to explain the top-quark forward-backward asymmetry.

We thank the Fermilab staff and the technical staffs of the
participating institutions for their vital contributions. This work
was supported by the U.S. Department of Energy and National Science
Foundation; the Italian Istituto Nazionale di Fisica Nucleare; the
Ministry of Education, Culture, Sports, Science and Technology of
Japan; the Natural Sciences and Engineering Research Council of
Canada; the National Science Council of the Republic of China; the
Swiss National Science Foundation; the A.P. Sloan Foundation; the
Bundesministerium f\"ur Bildung und Forschung, Germany; the Korean
World Class University Program, the National Research Foundation of
Korea; the Science and Technology Facilities Council and the Royal
Society, United Kingdom; the Russian Foundation for Basic Research;
the Ministerio de Ciencia e Innovaci\'{o}n, and Programa
Consolider-Ingenio 2010, Spain; the Slovak R\&D Agency; the Academy
of Finland; the Australian Research Council (ARC); and the EU community
Marie Curie Fellowship Contract No. 302103.

\bibliography{Afbb}

\begin{thebibliography}{35}
\bibitem [{geo()}]{geometry}%
  \BibitemOpen
  \bibinfo {note} {We use a cylindrical coordinate system with the origin at
  the center of the CDF II detector, $z$ pointing in the direction of the
  proton beam, $\theta$ and $\phi$ representing the polar and azimuthal angles,
  respectively, and pseudorapidity and rapidity defined by $\eta = - \ln
  \tan(\theta/2)$ and $y = (1/2) \ln (E + p_z / E - p_z)$, respectively. The
  transverse momentum $p_T$ (transverse energy $E_{T}$) is defined to be
  $p\sin\theta$ ($E\sin\theta$).}\BibitemShut {Stop}%
\bibitem [{Jet()}]{JetConeSize}%
  \BibitemOpen
  \href@noop {} {}\bibinfo {note} {{We use a fixed-size cone algorithm to
  identify jets with a cone radius $\Delta R = \sqrt{(\Delta \eta)^2 + (\Delta
  \phi)^2} = 0.7$.}}\BibitemShut {Stop}%
\end{thebibliography}
\bibliographystyle{apsrev4-1-JHEPfix}

\end{document}